%% file: paper.tex
\documentclass[runningheads]{llncs}
\usepackage{graphicx}
\usepackage{comment}
\usepackage{cite}
\usepackage{amsmath}
\usepackage{ wasysym }
\usepackage{ amssymb }
\usepackage{mathptmx}
\usepackage{amsmath}
\usepackage{pifont}

\usepackage{subfigure}
\usepackage{url}
\usepackage{color, wrapfig}
\usepackage{hyperref}
\usepackage{academicons}
\usepackage{xcolor}

\usepackage{times} 
\usepackage{amsmath} 
\usepackage{amssymb}  
\usepackage{enumitem}

\usepackage{wrapfig,lipsum,booktabs}

\begin{document}
\author{Abhinav Palia\inst{1}
\and
Rajat Tandon\inst{2}
\and
Carl Mathis\inst{1}} 
\authorrunning{A. Palia et al.}

\institute{Amazon Web Services (AWS), USA\\
\email{\{appalia,carmathi\}@amazon.com} \and
University of Southern California, USA\\
\email{rajattan@usc.edu}
}

\title{Utilizing Shannon's Entropy to Create Privacy Aware Architectures}

\maketitle   

\begin{abstract}

Privacy is an individual's choice to determine which personal details can be collected, used and shared.  Individual consent and transparency are the core tenets for earning customers’ trust and this motivates the organizations to adopt privacy enhancing practices while creating the systems.

\par The goal of a privacy-aware design is to protect information in a way that does not increase an adversary's existing knowledge about an individual beyond what is permissible. This becomes critical when these data elements can be linked with the wealth of auxiliary information available outside the system to identify an individual. Privacy regulations around the world provide directives to protect individual privacy but are generally complex and vague, making their translation into actionable and technical privacy-friendly architectures challenging. In this paper, we utilize \textit{Shannon's Entropy} (\textit{SE}) to create an objective metric that can help simplify the state-of-the-art Privacy Design Strategies proposed in the literature and aid our key technical design decisions to create privacy aware architectures.

\end{abstract}

\vspace{-4mm}

\input{intro}

\input{related}

\input{threatmodel}

\input{formulation}


\input{privacy-aware-design-strategies}

\input{discussion}


\input{conclusion}

\vspace{-4mm}
{\footnotesize \bibliographystyle{abbrv}
\bibliography{paper.bib}}

\end{document}

%% file: intro.tex
\vspace{-4mm}
\section{Introduction}

With the increasing awareness created by privacy community, individuals are realizing that it's their choice to determine what information can be collected, shared and processed by an organization. As a result, regulatory requirements are getting enforced and the organizations have started looking at different approaches to establish a privacy baseline for their customers. Among the approaches, Privacy-by-Design helps define privacy requirements at a high level. Also, there exists privacy enhancing technologies which are relevant after a system is developed~\cite{bluebook:2012}. For the regulations, most of the privacy laws are vague making it tough to translate them into technical solutions and further get complicated when organizations need to comply with multiple laws (for e.g., GDPR and PIPEDA)~\cite{hardcodeprivacy, confusion}. Although, frameworks like the NIST 800-53 v5~\cite{nist80053} and the NIST Privacy Framework~\cite{nistprivacy} are more prescriptive, they do not specifically cover privacy design strategies to help with privacy-aware technical architecture solutions. Notably, all these privacy laws and frameworks prohibit the identifiability of an individual.

\par Privacy-by-design framework focuses on embedding privacy into the design and business practices, and aims at achieving privacy-friendly architectures. In the literature, a number of design strategies are proposed but they do not provide any objective formulation that can support fundamental decisions for handling data at the architecture level. In particular, design strategies such as \textit{data minimization}, \textit{data separation}, \textit{data hiding}, \textit{data abstraction}, provide a generic guidance on how data can be stored to increase privacy of a system. But these strategies do not answer ``what" data should be minimized, and ``what" data attributes should be separated, hidden or abstracted~\cite{challengespbd}.

\par In this paper, we formulate \textit{Shannon's Entropy~(SE)} and apply it to the data oriented design strategies proposed by~\cite{bluebook:2012}. Our formulation attempts to refine and simplify these tactics, and provide a quantifiable measure to support key design decisions on what data attributes need to be handled for improving individual privacy in a system. The rest of the paper is organized as follows-- Section~\ref{Sec:Related} provides an overview of the related work. We define our threat model, formulation and privacy aware design strategies in Sections~\ref{Sec:ThreatModel},~\ref{Sec:Formulation} and~\ref{Sec:Design Strategies} respectively. In section~\ref{Sec:Discussion}, we discuss the applications of our formulation and the associated trade-offs, concluding our paper in section~\ref{Sec:Conclusion}.

%% file: related.tex
\vspace{-4mm}
\section{Related Work}
\label{Sec:Related}
Over the years, multiple surveys have been conducted~\cite{survey1, privacyepicsurvey}
which demonstrate customer's dissatisfaction with the organizations in data collection, processing and sharing. Because of the lack of or limited control over their own data, 
concerns have increased on how organizations are storing and handling all the customer data to prevent privacy leakage. Although a number of privacy regulations and standards have been enacted globally~\cite{privacylaws} to protect customer's right to privacy, these regulations do not provide any technical guidance on how systems should be built~\cite{hardcodeprivacy}. Privacy enhancing techniques~\cite{kanonymity, ldiversity} 
have been proposed in the literature but are useful only for a fully developed system and come either at the expense of utility of the system~\cite{tradeoff} or require rigorous optimization~\cite{abbygeo}. Privacy by design strategies and privacy aware architectural guidance~\cite{bluebook:2012} available in the literature, provide generic recommendations but there exists a gap in using these strategies objectively in real life systems.
In the past, \textit{SE} and Information theory concepts~\cite{noiseless, Shannon1} have been used to define privacy and create metrics to evaluate several privacy enhancing mechanisms~\cite{sankar2010} using mutual information but 
in this paper, we propose utilizing \textit{SE} as a quantifiable measure to define identifiability \& linkability which can assist objective decision making for creating privacy aware architectures and solutions 
while still maintaining the utility of the system.  

%% file: threatmodel.tex
\vspace{-4mm}
\section{Threat Model And Assumptions}
\label{Sec:ThreatModel}
The goal of threat modeling is to identify and enumerate potential threats to a system so that mechanisms can be implemented to prevent and avoid vulnerabilities. In this paper, we consider identification of an individual or \textit{identifiability} as our biggest threat if a database is breached or a system is compromised. This can help us prioritize and decide what privacy-aware architectural strategies can be used and how it can be implemented. Our threat model appears to be simpler than the LINDDUN model~\cite{LINDDUN-GO:2015} but we understand that identifiability is the ultimate risk as the consequence of \textit{linkability}, \textit{detectability}, \textit{unawareness} is also identifiability and inference. We believe that our proposed formulation is applicable to avoid privacy risks for \textit{disclosure of information} and in \textit{data sharing}. \textit{Non-Compliance} is out of scope for us as we treat identifiability as compliance agnostic and is a risk in all the regulatory frameworks. It has to be noted that our formulation focuses primarily on the \textit{unlinkability} goal out of the six "Privacy Protection Goals"~\cite{hansenprivacygoals:2015}. However, we use the remaining privacy goals as our basis to derive and optimize the privacy parameter using our formulation.

\textit{\textbf{Assumptions}}: For the sake of simplicity of our formulation, we start with the scenario that there exists a database in a system with one table containing records of individuals and the organization wants to implement a privacy-friendly architecture such that 
an adversary 
won't be able to identify or make inference about an individual of interest (\textit{IoI}) if the database is breached. Just for convenience, we will be using the terms database and table interchangeably. We will be focusing on relational databases which we assume can be extrapolated to the linkages that exist because of data being distributed in a number of data sources within a system. Lastly, we focus on the architectural tactics and assume that the only way attacker can interact with the system is by hacking into the server and not by asking queries to an online database.

%% file: formulation.tex
\vspace{-4mm}
\section{Formulation}
\label{Sec:Formulation}
\par\textbf{Identifiability:} We start our model formulation by defining identifiability. Identifiability of an individual $i$ means that the attacker can sufficiently identify an \textit{IoI} within a set of individuals. Identifiability is the opposite of anonymity and has one-to-one relationship with the attributes associated with $i$. Mathematically, if in a database, denoted by $D$, $\theta$ represents a set of attributes or characteristics that uniquely identify the individual $i$, and if $\theta_{i_1} = \theta_{i_2}$, then for identifiability, $i_1 = i_2$.

\par 
\textit{SE} quantifies the uncertainty of an event or the amount of information gained from an event. In simpler terms, more possibilities of an event lead to more uncertainty, and hence more \textit{information gain} when the event is revealed. Conversely, certainty of an event increases when there are less possible outcomes. If $H$ be the entropy, which is defined as the number of bits required to represent possible states or outcomes, then in order to have identifiability, $H = 0$, as $H = - p~ln(p)$, with $p=1$; where $p$ is probability of identifying an individual. 

 \par Let there be a database $D(A_1, A_2, \ldots, A_n)$, with $A_j$ denoting \textit{direct} identifiers or \textit{quasi}-identifiers. Examples of direct identifiers include SSN, email address, telephone number, or any other attribute that is unique over the distribution of attribute values (one-to-one and onto) \textit{globally}~\cite{investopedia}. In terms of entropy, if $X$ represents an event to identify \textit{IoI} $i$, given the knowledge of a direct identifier attribute value \{$A_{direct}(i)$\}, then the entropy equation for the event $X= i$, conditioned on the knowledge of direct identifier revealing $i$, $H(X = i | A_{direct}(i))$ = $0$. This equation holds for any combination of direct identifiers. Certain quasi-identifiers can still uniquely identify individuals~\cite{quasi-2011}. The entropy of an event $X = i$, given the distributions over $m$ quasi-identifiers ($A_{quasi\;  1}(i), A_{quasi\;2}(i)), \ldots A_{quasi\;m}(i)$) can be written as
\begin{center}
    $H(X=i|(A_{quasi\; k}))$ with $k= 1, 2,\ldots, m$ \quad $(1)$
\end{center}
Expanding using Bayes Rule,
\vspace{-2mm}
\begin{center}
    $H(X=i|(A_{quasi\; k})) = -p(X=i| (A_{quasi\;1}(i), A_{quasi\;2}(i), \ldots, A_{quasi\;m}(i))~ log[p(X=i| (A_{quasi\;1}(i), A_{quasi\;2}(i), \ldots, A_{quasi\;m}(i))];$
\end{center}
\vspace{-2mm}
\begin{center}
    = 
    $\frac{-p[(X=i, A_{quasi\;1}(i)=a_{q1}, A_{quasi\;2}(i)=a_{q2}, \ldots, A_{quasi\;m}(i)=a_{qm})]}{p(A_{quasi\;1}(i)=a_{q1}, A_{quasi\;2}(i)=a_{q2}, \ldots, A_{quasi\;m}(i)=a_{qm})}$~\\~\\$log\frac{p[(X=i, A_{quasi\;1}(i)=a_{q1}, A_{quasi\;2}(i)=a_{q2}, \ldots, A_{quasi\;m}(i)=a_{qm})]}{p(A_{quasi\;1}(i)=a_{q1}, A_{quasi\;2}(i)=a_{q2}, \ldots, A_{quasi\;m}(i)=a_{qm})}$ \quad $(2)$
\end{center}
where, $a_{qk}$ are the attribute values for individual $i$. Since for identifiability, $A_{quasi\; k}$ uniquely identifies $i$, we can write,
\begin{center}
     $ p(X=i, A_{quasi\;1}(i)=a_{q1}, A_{quasi\;2}(i)=a_{q2}, \ldots, A_{quasi\;m}(i)=a_{qm}) = p(A_{quasi\;1}(i)=a_{q1}, A_{quasi\;2}(i)=a_{q2}, \ldots, A_{quasi\;m}(i)=a_{qm})$.
     \\
\end{center}
Therefore, $(2)$ becomes, $H(X=i|(A_{quasi\; k})) = 1~ln(1) = 0$.

\par It is important to account for an adversary's knowledge of auxiliary information ($Aux.$) and the context ($C$). Needless to say, the auxiliary information or the context is not enough to satisfy the condition of identifiability of $IoI$ serving as the motivation for the attack. Context can be described as any information that can be used to characterize the situation of an entity, like an individual's, habits, emotions, or the metadata related to a situation. The organization storing individual's data has the right context but for the attacker, it can serve as an additional piece of information along with the auxiliary information. We would like to argue that the correctness of context can increase or decrease the knowledge of the attacker. If an attacker has the incorrect knowledge of context, it increases the uncertainty and conversely, it can augment the auxiliary information that the attacker possesses leading to identifiability. Using this, we can re-write $(1)$ as,
\begin{center}
    $H(X=i|(A_{quasi\; k}), Aux., C)$ \quad $(3)$
\end{center}
This leads us into defining \textit{Individual Privacy Parameter}~(IPP) represented by $\varepsilon$. For deriving the privacy equation, we propose that an adversary's knowledge of an $IoI$ should not increase beyond what is already known to him if the database is breached, giving us an upper bound for $H$. Therefore,
\begin{center}
    $0 \leq H(X=i|(A_{quasi\; k}), Aux., C) \leq H(X=i| Aux., C)$, 
    \\~\\
    $0 \leq \varepsilon \leq 1 $, 
      $\varepsilon =\frac{H(X=i|(A_{quasi\; k}), Aux., C)}{H(X=i| Aux., C)}$ \quad $(4)$
\end{center}
IPP $\epsilon$ represents the current state privacy of the individual records in the database based on the distribution of attributes associated with them. Differential privacy~\cite{DP_cynthia:2006}, which is considered as the strongest definition of privacy, defined for a mechanism $M$ on a database $D$ as, $P[M(D) \in Y] <= e^{-\varepsilon}.(P[M(D') \in Y])$; with $|D-D'| = 1 $.

\par Comparing IPP-$\varepsilon$ with differential privacy $\varepsilon$ (DP-$\varepsilon$), adding noise to the database increases the entropy and protects identifiability.~\cite{noiseless} followed a similar approach to exploit the entropy already present in the database and substitute that for external noise to the output. However, addition of noise affects the utility of the system which is not the case with IPP-$\varepsilon$. Further, unlike DP-$\varepsilon$, the basis of optimal value of IPP-$\varepsilon$ is individual's consent and we discuss this in detail in our next section.

\par\textbf{Linkability:}
Linkability, $\lambda$, between two or more items within a system means that the knowledge of the system can reveal the relation, denoted by $\backsim$, between the items which was not known to the adversary with their prior knowledge or auxiliary information, $Aux.$~\cite{terminology}. The more commonly used terminology is unlinkability described from an adversary's perspective, specifying the before and after states of an adversary observing the system which can be quantified as perfect preservation of unlinkability 
~\cite{terminology}. As pointed out earlier, the ultimate privacy harm associated with linkability is that it can lead to identifiability if too much linkable information is combined or can lead to a potential inference if a link to a sensitive attribute is revealed~\cite{LINDDUN-GO:2015}. 
\par
\begin{wrapfigure}{l}{0.45\textwidth}
\vspace{-9mm}
 \begin{center}
    \includegraphics[width=0.45\textwidth]{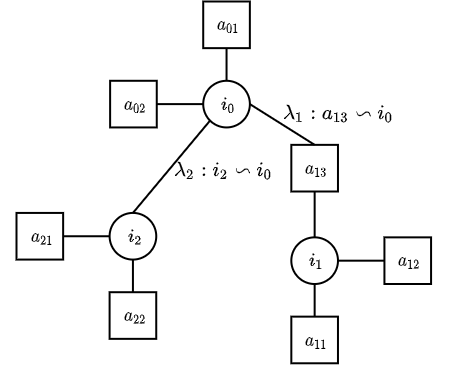}
  \end{center}
  \vspace{-6mm}
  \caption{Possible linkages for \textit{IoI} $i_0$.\label{fig:links}}
  \vspace{-6mm}
\end{wrapfigure}

 We start formulating linkability for our database model by examining the possible linkages for \textit{IoI} $i$ under the notion `is related to' $i$ denoted by $\backsim_{(i)}$ similar to the definition in~\cite{linkability}.
In a relational database, as presented in Figure \ref{fig:links} for \textit{IoI} $i_0$, the two linkages are $\lambda_1: a_{13}\backsim i_0$ and $\lambda_2: i_{2}\backsim i_0$. $\lambda_1$ represents another individual's ($i_1$) attribute `is related to' $i_0$ and $\lambda_2$ denotes another individual ($i_2$) `is related to' $i_0$. 
For the first case, $a_{13}$ can be treated as another quasi-identifier ($A_{{quasi}-{linked}}$) which can be plugged into eq. (3) above ultimately leading to identifiability. For $\lambda_2$, if the relation between some individual and \textit{IoI} $i_0$ is revealed to an adversary, which is very common in online social networks~\cite{sociallink1},
 it can lead to inferences about the attribute \textit{values} associated with $i_0$. We propose that the knowledge of the nature (an attribute or auxiliary information) and the context of the relation is a necessary but \textit{not} a sufficient condition to draw a relevant inference about an individual or to reinforce an inference leading to identification. For the sake of simplicity of discussion, we will scope ourselves to $\lambda_1$ in this paper as $\lambda_2$ relations are also actualized via underlying attributes. 


%% file: privacy-aware-design-strategies.tex
\vspace{-4mm}
\section{Privacy Aware Design Strategies}
\label{Sec:Design Strategies}
%

Privacy regulations around the world are generally non-technical~\cite{solove2008understanding}, and the translation of these laws into design solutions and architectures is challenging. Privacy enhancing techniques are suitable for systems that are already developed, complex to implement if the organization is not mature in the area of privacy and most of these techniques come at the cost of utility of the system.
\par In this paper, we explore Hoepman’s eight privacy design strategies~\cite{bluebook:2012} to build privacy friendly systems. Privacy is an individual's choice since one is the owner of one's information. In data processing, transparency is vital and one of the most important goal that an organization needs to accomplish to earn their customer's trust~\cite{trust1}.
Out of the eight design strategies listed in the blue book of privacy~\cite{bluebook:2012}, four strategies are process-oriented, focusing on the procedural aspects of data handling in an organization. Individuals should be informed about ``what", ``why" and ``when" the data is collected, stored, processed, or shared by an organization and explainable mechanisms~\cite{explainable_ethics:2020} should be developed which can help demonstrate privacy awareness, establishing trust and increasing customer's confidence. The core tenet for our formulation presented above and architectural discussions in this section is the individual's control and consent which can help derive an optimum value of IPP ($\varepsilon_0$). It is the organization's responsibility to preserve every individual's privacy and prevent identifiability. Ideally, $0 < \varepsilon_0 \leq \varepsilon$, where $\varepsilon$ is the target IPP value calculated for a user in the database. Next, we examine the data oriented strategies in detail utilizing our formulation to help us create a privacy friendly architecture. 

\par\textbf{Data Hiding}: Based on our formulation, all the direct identifiers in the database $A_{direct}$, whose knowledge makes $H(X=i)=0$, should be hidden and stored separately from other direct and quasi-identifiers. The access to these tables containing direct identifiers must be restricted by utilizing authentication and authorization mechanisms, and data should be protected by obfuscation, anonymization, or encryption.
\begin{wraptable}{r}{7cm}
\centering
\vspace{-8mm}
    \begin{tabular}{ l }
    
    \hline
    \textit{find-risky-comb.}($(A_{quasi\; k}(i)), \varepsilon_0$) \\ 
    \hline
    \\
    $T$: Tabulation Table, \\
    \textit{Risky-Attribute-Set($i)$}: Set of risky attributes for\\individual $i$;
    \\
 
    For every combination of $A_{quasi\; k}(i)$ \\ \quad DP-Tabulation($temp\_\varepsilon$, \{$A_{quasi\; k}$\}) $\leftarrow T$ \\
    For every cell in $T$ \\ \quad \textit{if}$(temp\_\varepsilon \leq \varepsilon_0)$: \\
    \quad \quad  \{$(A_{quasi\; l})$\} $\leftarrow$ \textit{Risky-Attribute-Set($i$)}\\
    \\
    \hline
    \end{tabular}
    \caption{Function to identify risky combinations of quasi-identifiers}
\label{table:risky-comb}
\vspace{-8mm}
\end{wraptable} We propose using a \textit{local} identifier replacing the direct identifiers everywhere else in the system for referencing that individual while making it harder for an attacker to identify an individual using a direct identifier if the database is breached.

\par\textbf{Data Separation}: It is crucial to logically and physically separate personal data into multiple databases and on different servers to avoid linkability and prevent identification of an individual~\cite{dataseparation}. 
It is important to design database schema in such a way that the \textit{risky} attribute combinations are separated which could otherwise lead to identification of linkages and ultimately identifiability. We propose the function \textit{find-risky-comb.}() in Table I, that utilizes our formulation from section \ref{Sec:Formulation} and dynamic programming (\textit{tabulation}) to identify the risky combinations of quasi-identifiers.

\par The outcome of \textit{find-risky-comb.}() can help make an informed decision on what attribute values can be stored in a table to guarantee individual privacy. We can use multiple strategies such as replacing the data values for the risky attributes by a pointer to a separate table protected by security mechanisms where the actual values can be stored. We will discuss the performance trade-offs in Section \ref{Sec:Discussion}.




\par\textit{\textbf{Data Minimization}}: 
Every data element can potentially contribute to the identification of an individual. Therefore, it is safe to say minimizing data can help prevent identifiability and avoid linkability. In order to minimize data, \\(i) List all the attributes required for the current functionality of the application (data flow map and inventory).
\\(ii) Check if all these attributes have individual's consent and identify the difference.
\\(iii) For the attributes without user consent, re-evaluate their use in the current state and future state of the application and inform the individual giving them the alternative of opting out. 
\\(iv) Strip, or destroy all the attribute values for which the customer has not consented for current or future use.
   
\par \textit{\textbf{Data Abstraction}}: Based on our formulation, adding \textit{similar} records (same values for quasi-identifiers) in the database or perturbing the database decreases the probability of identifying an individual. Other techniques such as splitting the attributes to perturb the context, grouping of information for query based system, storing summarized or generalized attributes, instead of granular data for every individual should be used since $H(X=i|(A_{identifier}))$ is dependent on the uniqueness of the combination of attributes to identify $i$.


%% file: discussion.tex
\vspace{-4mm}
\section{Discussion}
\label{Sec:Discussion}
In this section we discuss the application of the design strategies using our formulation, and the trade-offs involved. 
The greatest advantage of utilizing these design strategies is that it gives users the control over their data, without affecting the query outcome or utility of the system. Our formulation takes user consent and context into consideration which can help organizations fulfill some of the compliance requirements. We understand the dependency for some of the strategies such as Data Separation is the universe of data for our formulation to compute the risky combinations. However, over a period of time, machine learning can be used to develop organization specific database schema patterns and groupings for similar $\varepsilon$ values generalizing the risky attributes. Another limitation we note is the overall performance of the queries if the data is distributed across different databases on multiple servers. We propose using logical design strategies such as leveraging microservices architecture~\cite{microServices}
and aggregating them based on their interactions with the database which can help reduce the performance impact of data separation. And lastly, another application of the formulation and design strategies we propose is to create sanitized data sets for sharing it with third parties by eliminating the risks of linkability and identifiability. 

%% file: conclusion.tex
\vspace{-4mm}
\section{Conclusions}
\label{Sec:Conclusion}
In this paper, we use \textit{SE} to create an objective measure of privacy to understand and help us take key design decisions around data attribute storage in databases in order to achieve privacy friendly architectures. We focus on identifiability of individual as our biggest threat and detail how we can determine risky combinations of quasi-identifiers which can lead to individual identification and further, how we can use the design strategies proposed in literature rationally and objectively using our formulation. In our future work, we will demonstrate the model proposed in this paper by building an automated system which can create a dashboard for the system architects helping them identify the risky combinations of attributes and provide relevant privacy recommendations along with their performance implications which can be used as a basis to create flexible privacy aware architectures for live systems.